\documentclass[5p,authoryear]{elsarticle}
\usepackage[utf8]{inputenc}
\usepackage[T1]{fontenc}
\usepackage{textcomp}
\usepackage{amsmath}
\usepackage{graphicx}
\usepackage{caption}
\usepackage{subcaption}
\usepackage[hidelinks]{hyperref}
\usepackage{cleveref}
\usepackage{booktabs}
\usepackage{amssymb}

\usepackage{tabularx}
\usepackage{multirow}


\begin{document}

\title{A Multimodal Framework for Topic Propagation Classification in Social Networks}
\author{Yuchuan Jiang}
\author{Chaolong Jia\corref{cor1}}
\author{Yunyi Qin}
\author{Wei Cai}
\author{Yongsen Qian}

\address{Chongqing University of Posts and Telecommunications, School of Software Engineering, Chongqing, China}
\cortext[cor1]{Corresponding author:jcl@cqupt.edu.cn}

\begin{abstract}
   The rapid proliferation of the Internet and the widespread adoption of social networks have significantly accelerated the speed of information dissemination. 
   However, this transformation has introduced new complexities in information capture and processing, presenting substantial challenges for researchers and practitioners. 
   Consequently, predicting the dissemination of topic-related information within social networks has become a critical focus in contemporary research. 
   This paper presents a predictive model for topic dissemination in social networks, integrating multidimensional features derived from key characteristics of the dissemination process. 
   First, this paper introduces two novel indicators—user relationship breadth and user authority—into the PageRank algorithm, quantifying these factors to facilitate a more effective assessment of user influence. 
   Second, the Text-CNN model is employed for sentiment classification, extracting sentiment features from text content analysis. 
   Next, temporal embeddings of the nodes are encoded using the Bi-LSTM model, capturing the temporal dynamics of these nodes. 
   Additionally, the paper captures user interaction traces with topics over time, replacing the traditional metric of topic views with a more precise measure of communication characteristics. 
   Finally, the multidimensional features are fused using the Transformer model, significantly enhancing the model's predictive power. Experimental results demonstrate that the proposed model outperforms traditional machine learning and unimodal deep learning models regarding FI-Score, AUC, and Recall, validating its effectiveness and superiority in predicting topic propagation within social networks.\newline

\end{abstract}

\maketitle

\section{Introduction}
With the advent of the big data era, an increasing number of users are expressing their personal opinions and views on social network platforms. The prediction of information dissemination within social networks spans multiple research domains, including events, hashtags \citep{Weng2015,Zhu2003}, and topics. Among these, hot topics—key nodes that rapidly capture users' attention —play a pivotal role in the topic propagation process\citep{Eytan2011}. Research has demonstrated that the spread of topic information is not random \citep{Zhao2015}, but instead follows discernible spreading patterns \citep{Shard2012}. A comprehensive understanding of these patterns and the ability to predict them effectively can provide valuable insights into the dynamics of social opinion, thereby contributing to the maintenance of societal stability and security.

Therefore, understanding the principles of social network communication requires an in-depth analysis of key characteristics, such as user attributes, dynamic interactions, and communication time sequences. Examining these factors enables a fundamental analysis of information dissemination, facilitating deeper insights into social networks.

However, most existing information dissemination prediction methods predominantly rely on a single dissemination sequence or user feature, lacking comprehensive models that fuse multidimensional features. As a result, the prediction of information dissemination through multi-feature fusion remains a critical and under-explored research direction. Current research faces several key challenges: 1) Many topic prediction methods, particularly those based on LSTM and its variants, fail to adequately explore topic relevance and suffer from the limitation of not supporting parallel training, leading to reduced processing efficiency; 2) Existing studies primarily focus on user and propagation features, with insufficient attention given to temporal features during propagation, and a lack of effective integration of temporal features with other types of features; 3) In the feature extraction process, hidden emotional features of users are often overlooked, and the assessment of user influence typically relies on a single indicator, which diminishes the accuracy and authenticity of topic propagation predictions.

To address these challenges, this Paper is guided by the necessity of multidimensional feature fusion, the computational advantages of Transformer models, and the influence of temporal and emotional dynamics in social networks. First, the inefficiency of parallel computation in LSTM-based models motivated the adoption of Transformer networks, which leverage self-attention mechanisms to capture long-range dependencies and enable efficient parallel processing. Second, underutilizing temporal features in propagation prediction led to integrating Bi-LSTM to extract sequential dependencies, ensuring a more comprehensive temporal representation. Third, the increasing recognition of emotional influence in social media dissemination inspired the incorporation of Text-CNN for sentiment analysis, enabling the model to capture hidden emotional cues that drive topic propagation. Finally, to overcome the limitations of single-metric user influence assessment, a PageRank-based authority evaluation was introduced to dynamically quantify user impact by considering both propagation behavior and social structure.The rough process is shown in \textbf{Fig. 1}, which demonstrates the abbreviated ideas obtained from the information structure as well as the nature of social networks in this study.

\begin{figure}[h]
   \centering
   %\hspace{-5.5cm}
   \includegraphics[height=7cm,width=8.8cm,scale=3]{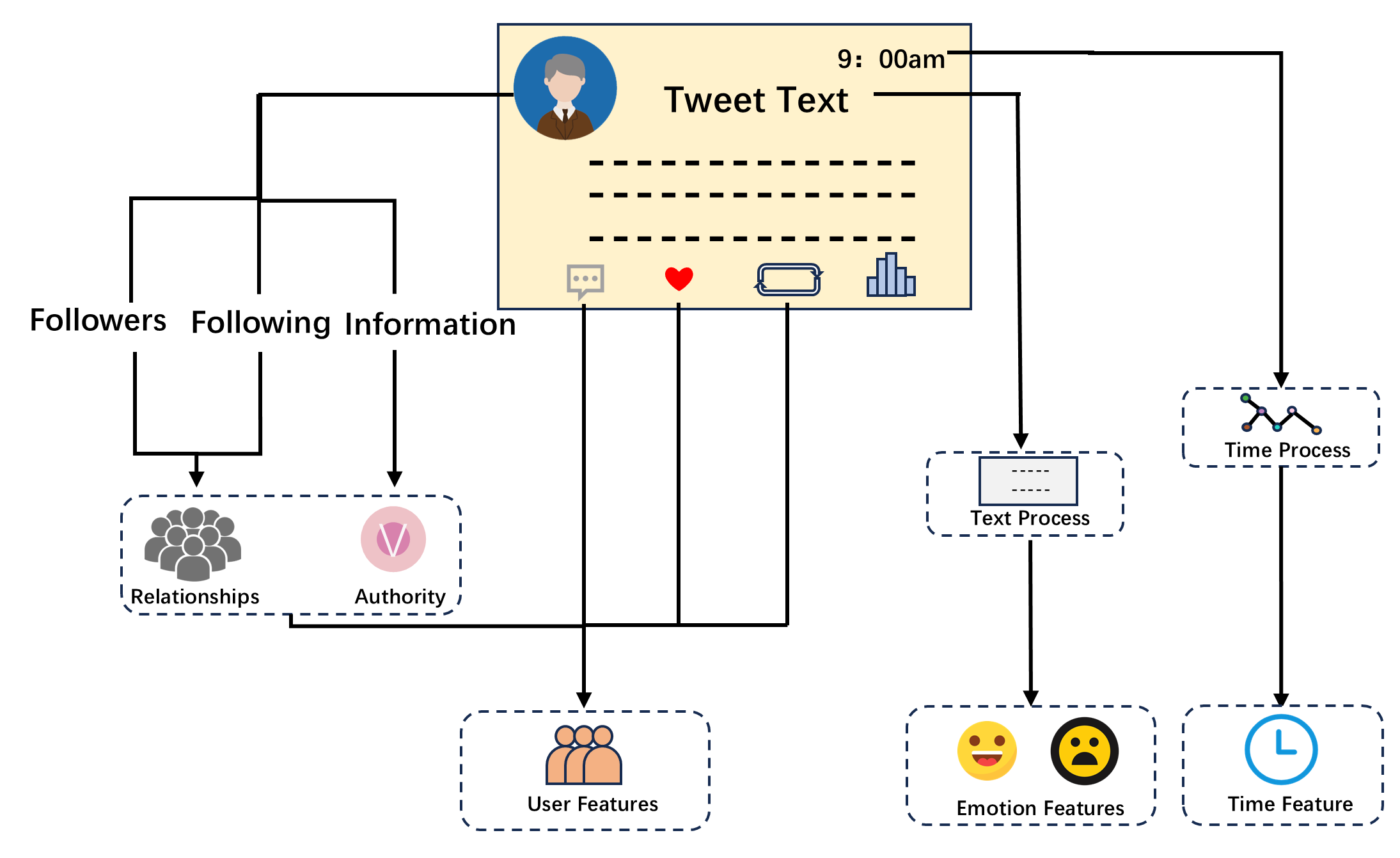}
   \caption*{Fig.1.Idea Inspiration Process Diagram}
\end{figure}

Consequently, this paper proposes a series of enhancements: For Problem 1, the Transformer model is introduced, and multidimensional features in the propagation process are integrated, with careful consideration of topic correlation and the complexity of the propagation process. Additionally, multi-threaded parallel processing is employed to improve computational efficiency. For Problem 2, temporal features of nodes are extracted using the Bidirectional Long Short-Term Memory (Bi-LSTM) network, and a novel feature fusion method is proposed to combine temporal features, user features, emotional features, and propagation features as Query, Key, and Value inputs to the Transformer model, significantly improving the fusion effect. For Problem 3, the Text-CNN model is utilized to extract emotional features from the text, and dynamic social behavior and user authority are incorporated to assess user influence more comprehensively through the PageRank algorithm.

The main contributions of this paper are as follows:

1) Introducing the Text-CNN model to conduct an in-depth analysis of topic text content, significantly enhancing the accuracy of sentiment analysis.

2) Replacing multi-dimensional heterogeneous graph  s with one-dimensional features, conforming to the propagation law through serialization, extracting temporal features using the Bi-LSTM model to solve the long-term dependency problem, and improving the processing capability of temporal data to enhance prediction accuracy.

3) Proposing a novel feature fusion method to input multidimensional features into the decoder layer of the Transformer, thereby creating a topic information propagation prediction method based on multi-feature fusion, which effectively improves the authenticity and accuracy of predictions.

The subsequent sections of this paper are organized as follows: Section 2 reviews current research on topic dissemination in detail; The section 3 defines key materials related to social network topic data, providing a foundational understanding to enhance network security stability; Section 4 outlines the primary methods of this research, including the integration of new indicators with traditional features, the implementation of multi-feature fusion, and the application of the Transformer model in multidimensional feature fusion. Section 5 presents experimental results, validating the effectiveness of the proposed method through metrics such as FI-Score and Precision, along with model comparisons and ablation studies.Section 6 examines the importance of topic information categorization in social networks for monitoring public opinion and safeguarding online security, integrating insights from the experimental results presented in Section 5. Finally, Section 7 summarizes the research findings and discusses future research directions.

\section{Related Work}
In recent years, numerous crises fueled by public opinion have occurred, prompting increasing scholarly attention to the study of public opinion topics. Predicting the propagation of public opinion on a given topic can facilitate more effective control of public sentiment, thereby mitigating the potential societal harm caused by such events.

Currently, a substantial body of research exists on topic propagation prediction, with one of the most prominent approaches being the application of the Susce  ptible-Infected-Recovered (SIR) model and its variants \citep{YangandZhang2022,Li2023,Wang2014,Zhu2022}. This approach analogizes topic propagation to the viral spread of infectious diseases \citep{Romualdo2015}, aiming to uncover topic propagation patterns and facilitate accurate predictions. Additionally, numerous studies have employed Convolutional Neural Networks (CNNs) and their variants, leveraging deep learning techniques to model topic propagation, particularly by treating social network topics as time-series data \citep{Zheng2014}. Other research has utilized Graph Convolutional Networks (GCNs) to model topic information propagation through the network topology and to forecast the propagation trends of subsequent topics \citep{Petar2018,Chen2021}.

All in all, research on topic communication prediction can be broadly classified into three main approaches: feature extraction-based methods, model-bas  ed methods, and deep learning-based methods.

Feature extraction-based methods rely heavily on expert knowledge for manually selecting and extracting features from the propagation process, often based on prior assumptions. This approach introduces subjectivity, leading to uncertainty in the selected features, which can undermine their generalization ability. For example, Zhang Mingjie et al. \citep{Zhang2023} proposed a multi-feature fusion method for microblog information dissemination prediction based on XGBoost, which enhances prediction accuracy by incorporating various features. Similarly, Zhian Yang et al. \citep{Yang2024} introduced G-Informer, a fusion model utilizing a graph-attention mechanism based on historical sequences, demonstrating robust performance with long-time sequences. Kyoungsoo Bok et al. \citep{Kyoungsoo2021} improved the TF-IDF algorithm by proposing a hot-topic prediction method incorporating user influence and expertise, leading to better recall and precision. Shulin Cheng et al. \citep{Cheng2022}introduced MUI, a microblogging user influence algorithm, which optimizes information dissemination by analyzing user interest similarity and dissemination ability. Jingqi Wang et al. \citep{Li2022} developed the TPP-DA model, which integrates user characteristics and topic influence to enhance prediction accuracy and truthfulness. Korolev et al. \citep{Korolev2018} utilized orthogonal non-negative matrix factorization (ONMF) for user topic analysis, proposing a time-series prediction method for user identification. Fang Ze-Han et al. \citep{FangandChen2022} introduced a crowdsourcing intelligence-based feature selection and collaborative trend prediction method, significantly improving the prediction of public opinion and communication of popular topics. Liu Yunxin et al. \citep{Liu2020} combined the K-means classification method with gating units and attention mechanisms to predict complex univariate time series, reducing sequence noise and improving prediction accuracy. Feroz et al. \citep{Feroz2023} studied social network topology from the perspective of investors, providing insights into network size and structure.
These studies introduce a variety of effective feature extraction and propagation prediction methods bas  ed on feature extraction approaches, achieving progress in improving propagation prediction accuracy. However, challenges remain in terms of the subjectivity of feature selection, model generalization ability, and the handling of dynamic changes and noise. Future research should focus on improving the robustness and adaptability of these models, as well as enhancing computational efficiency when processing large-scale social network data.While feature extraction-based methods have contributed to the advancement of information dissemination prediction by introducing diverse features and improving algorithms, they are still limited by several shortcomings. Notably, the subjectivity and uncertainty inherent in the feature selection process present significant challenges. In particular, for dynamically changing data, it is difficult to ensure that the extracted features generalize well across different contexts. Thus, future research should prioritize optimizing the feature extraction process, reducing subjectivity, improving mod  el stability under dynamic conditions, and increasing computational efficiency when working with large-scale datas  ets. This will be essential to further enhancing the accuracy and applicability of information dissemination prediction models.

Model-based approaches, in contrast to feature extraction methods, embed the core dynamics of the information dissemination process directly into predictive models. These models are typically extensions of contagion models, which simulate the spread of information as a dynamic process. For example, Shang Hongyun et al. \citep{Shang2016} proposed a Markov model based on the content state of a topic, which accurately and efficiently predicts the short-term popularity trend of a topic. Similarly, Yang Yi-Heng et al. \citep{Yang2023} enhanced the classic SIR model by introducing a modified contagion factor, improving the model's generalization ability for predicting information spread. Zhang Yuexia et al. \citep{ZhangandPan2021} proposed an improved hierarchical SIRS model, which more accurately captures the transmission dynamics of information across networks.

Other studies have integrated additional data sour  ces to enhance prediction accuracy. For instance, Keyang Ding et al. \citep{Ding2024} developed the Chinese Social Sentiment Prediction (LCSEP) dataset and proposed the HTEAM model for sentiment prediction, which successfully captured the social sentiments triggered by hot topics. Liwei Xu et al. \citep{Xu2023} combined text analytics with sentiment computation, and based on the grey prediction model, introduced a novel method for trend prediction, yielding better results in online public opinion forecasting. Rui Li et al. \citep{Li2022} utilized geographic case-based reasoning to predict public opinion trends on urban planning by integrating city-specific cases. Hengmin Zhu et al. \citep{Zhu2022} proposed a decision tree model for online topic hotness prediction, incorporating features such as publisher information and external content, which enhanced the prediction accuracy. Tianji Dai et al. \citep{Dai2022} used an improved SVM model to predict user retweet behaviors, achieving better topic prediction accuracy. Zhuomin Chen et al. \citep{Chen2023} introduced the Topic and Structure of Sensing Neural Network (TSNN) for rumor prediction, significantly improving prediction reliability. Additionally, Park, Eunhye, et al. \citep{Eunhye2021} applied the SARIMA model and topic variables to travel news data, successfully predicting travel demand.
These model-based approaches have effectively predicted topic popularity and sentiment, but still face limitations. Most contagion models assume linear and homogeneous dissemination, neglecting behavioral differences and dynamic network changes. Additionally, while sentiment and text analysis capture content features, existing models struggle with complex sentiment shifts and dynamic contexts. Future work should focus on refining these models to account for social network dynamics and heterogeneity, improving prediction accuracy and robustness.

Deep learning-based methods have become the mainstream approach for public opinion prediction due to their strong generalization capabilities and ability to automatically extract features, reducing human errors. For instance, Huang Yinggang \citep{Huang2021} proposed a hybrid LSTM-CNN model to reduce feature homogenization and improve offline prediction accuracy. Fan Wei et al. \citep{FanandLiu2022} developed a Transformer-based model, SST, that incorporates temporal and structural features to better capture information relevance. Xuegang Chen et al. \citep{Chen2023b} used gray prediction and fuzzy neural networks to analyze emergency impacts on public opinion, showing better handling of emergencies. Yuanyuan Zeng et al. \citep{ZengandXiang2023} introduced Persistence Enhanced Graph Convolutional Networks (PT-GCNs), combining propagation features with persistent content for topic prediction. Other notable contributions include Ranran Wang et al. \citep{Wang2023} with PT-GCN and Qiumeng Chen et al. \citep{ChenandShen2022}, who used large-scale graph neural networks for predicting information diffusion. Wenya et al. \citep{Wen2023} combined time-evolving graph neural networks with gated recurrent units to predict opinion hotspots in time series data.
These deep learning approaches have demonstrated strong performance in multi-dimensional feature modeling and generalization. However, challenges remain, particularly when modeling complex dynamic processes. Existing models often overlook the impact of spatio-temporal relationships, especially in large-scale social networks, where temporal and structural dynamics may not be adequately captured. Moreover, the interpretability and inference efficiency of deep learning models continue to be significant challenges. Future work should focus on optimizing spatiotemporal feature extraction, enhancing multi-level and cross-domain data fusion, and improving model interpretability and computational efficiency to better support real-world decision-making and strategy adjustments.

\section{Model Definitions and Materials}
\subsection{User Self-influence}
A social network is a complex, user-centered structure in which users, as communication nodes, play a pivotal role in the dissemination of information. In particular, key nodes—due to their unique characteristics—tend to have a broader influence and maintain closer associations with other nodes. As a result, studying the impact of key nodes on the characteristics of topics is of significant importance. The central focus of key node research lies in evaluating and analyzing their influence within the network.

User self-influence primarily quantifies the degree to which a node influences other nodes within a network. It is derived from fundamental user data and incorporates several factors, including industry market size, user activity, and the average development level of the industry. As a comprehensive metric, user self-influence offers a more objective reflection of user dynamics in online communities and their role in information dissemination compared to other reference indicators.

This paper introduces novel influence factors—user relationship breadth and user authority—as key components for measuring user self-influence. 
\subsubsection{User relationship breadth}
User relationship breadth, which quantifies the social connections between users, serves as a key indicator of a user’s social influence. In the topology of a social network, a node’s influence is directly related to the number of its connections: the more connections a node has, the broader its information dissemination capacity and, consequently, its influence. Thus, user relationship breadth can be represented by the number of links associated with a node. The larger the breadth, the more nodes the user can influence, thereby amplifying the user’s overall influence within the network.
By quantifying user relationship breadth, the influence of a user can be effectively evaluated. The formula for calculating user relationship breadth is provided in Eq.1:
  \begin{equation}\label{eq1}
    PR_{people}=\frac{N_i}{N_{\max}}
  \end{equation}
  where $PR_{people}$  represents the user relationship breadth, ${N_i}$ denotes the advocates of user $i$, and $N_{\max}$ indicates the maximum number of advocates associated with the selected user.
\subsubsection{User authority}
User authority is widely regarded as a key indicator of the influence of content posted by users. During the process of topic dissemination, users with higher authority tend to exert greater influence. User authority is determined by both user behavior and user recognition. Specifically, user behavior reflects an individual’s actions, which directly contribute to their perceived authority, while user recognition represents the extent to which other users acknowledge and validate this authority.
In this context, this paper introduces a novel indicator for measuring user authority, which considers both user behavior and user recognition. The calculation formulas for this indicator are presented in Eq. (2) and Eq. (3).
 \begin{equation}\label{eq2}
   UAD\left( U_i \right) =\alpha B_{verified}+\beta \frac{NR_{tweet}}{NR_{\max}}
 \end{equation}

 \begin{equation}\label{eq3}
   B_{verified}\left\{ \begin{array}{l}
      1\text{,User is not authenticated}\\
      1.5\text{,User is authenticated}\\
   \end{array} \right. 
 \end{equation}

 Where $UAD\left( U_i \right)$ represents the authority score of user.   $NR_{tweet}$ denotes the number of retweets sent by the user, while $NR_{\max}$ corresponds to the maximum number of   retweets among users within the dataset, and $B_{verified}$ indicates whether the user account is verified.Where $\alpha \text{,}\beta$ is an optional parameter, determined by specific requirements.
\subsubsection{Self-impact indicators}
In this paper, we introduce user relationship breadth and user authority as new self-influence metrics, and consider user self-influence measurement factors as a linear combination of the two metrics.The specific formula is shown in Eq. (4).

\begin{equation}\label{eq4}
   PR_{influence}\left( U_i \right) =PR_{people}+UAD\left( U_i \right) 
\end{equation}

Where $PR_{influence}\left( U_i \right)$ represents the user's self-influence indicator, $PR_{people}$ denotes the user's relationship breadth, and $UAD\left( U_i \right)$ corresponds the user's authority degree

\subsection{User Browsing Trace}
The propagation characteristics of a topic are often associated with the number of views its related articles receive. However, view counts are frequently inaccessible, and even when available, they can be skewed by repetitive user activity—i.e., a single user repeatedly viewing the same content. This redundancy undermines the reliability of view counts as an accurate measure of propagation dynamics.

To address these limitations, this study employs user browsing traces as an alternative to article view counts, leveraging interaction data to capture the nuances of topic dissemination more effectively. Specifically, the propagation characteristics of topics are analyzed through three key dimensions: topic resonance, which reflects the depth of audience engagement; direct propagation factor, which quantifies the immediacy and reach of direct interactions; and topic exposure, which measures the breadth of content visibility across the audience. This multidimensional framework offers a clearer and more robust representation of topic spread dynamics.

\subsubsection{Topic resonance}
Topic resonance,which is the number of likes, quantifies the level of interaction and engagement between users and a given topic. In this study, topic resonance is a key metric for analyzing user browsing traces, capturing how users recognize and approve of a topic's content. A higher resonance value reflects a greater number of users positively engaging with the topic, thereby enhancing the likelihood of its dissemination. Consequently, tracking topic resonance over a defined period provides valuable insights into the dynamics and trends of topic propagation. The computation of topic resonance is formally expressed in Eq. (5).

\begin{equation}\label{eq5}
  AP\left( T_i \right) =\sum_{j=1}^n{N_{Aj}\left( T_i \right)}
\end{equation}

$AP\left( T_i \right)$ denotes the total resonance of the topic, and $N_{Aj}\left( T_i \right)$ refers to the number of likes of $j$ at different moments of the topic $T_i$ $N_{Aj}$, which can be computed by counting the number of likes of different texts under the topic to get the total resonance of the topic as one of the propagation characteristics of the topic.

\subsubsection{Direct transmission factor}
When users engage with topics, they often exhibit direct interaction behaviors, such as retweeting related content. The total number of text retweets within a topic serves as a direct indicator of user contributions to its dissemination. By aggregating the number of retweets, it is possible to effectively estimate the depth of topic propagation. This metric, referred to as the direct dissemination factor, provides a quantitative measure of the dissemination characteristics of the topic. The computation is formally defined in Eq. (6).

\begin{equation}\label{eq6}
   TP\left( T_i \right) =\sum_{j=1}^n{N_{Tj}\left( T_i \right)}
\end{equation}

$TP\left( T_i \right) $ denotes the direct propagation factor of the topic, and $N_{Tj}\left( Ti \right)$ denotes the number of individual retweets of topic $T_i$ at different moments of $j$ $N_{Tj}$,By counting the number of retweets of the topic in different time periods, we get the total number of retweets of the topic as the direct propagation factor of the topic, and thus we get the topic'sPropagation Characteristics.

\subsubsection{Topic Exposure}
Topic exposure reflects the extent to which users are aware of and engaged with a topic. While users may casually browse a topic without delving into its content, commenting on a topic demonstrates a higher level of engagement, indicating that the user has read and understood the material more thoroughly. Consequently, the number of comments serves as a more precise indicator of topic exposure, effectively capturing the depth of user interaction and interest.

Therefore, this paper introduces the total number of topic comments as a new measure to replace the past view count to measure the topic exposure, which can substantially improve the prediction accuracy of the topic spread probability.The calculation formula is shown in Eq. (7).

\begin{equation}\label{eq7}
   CT\left( T_i \right) =\sum_{j=1}^n{N_{Cj}\left( T_i \right)}
\end{equation}

$CT\left( T_i \right)$ shows the total exposure of topic $T_i$, and $N_{Cj}\left( T_i \right)$ denotes the number of individual comments $N_{Cj}$ at different moments $j$ of topic $T_i$.By counting the number of topic comments at different times, the total number of topic comments is obtained, and finally the propagation characteristics of the topic are obtained.

The process of extracting user features is illustrated in \textbf{Fig. 2}.

\begin{figure}[h]
   \centering
   %\hspace{-5.5cm}
   \includegraphics[height=8cm,width=5.5cm,scale=3]{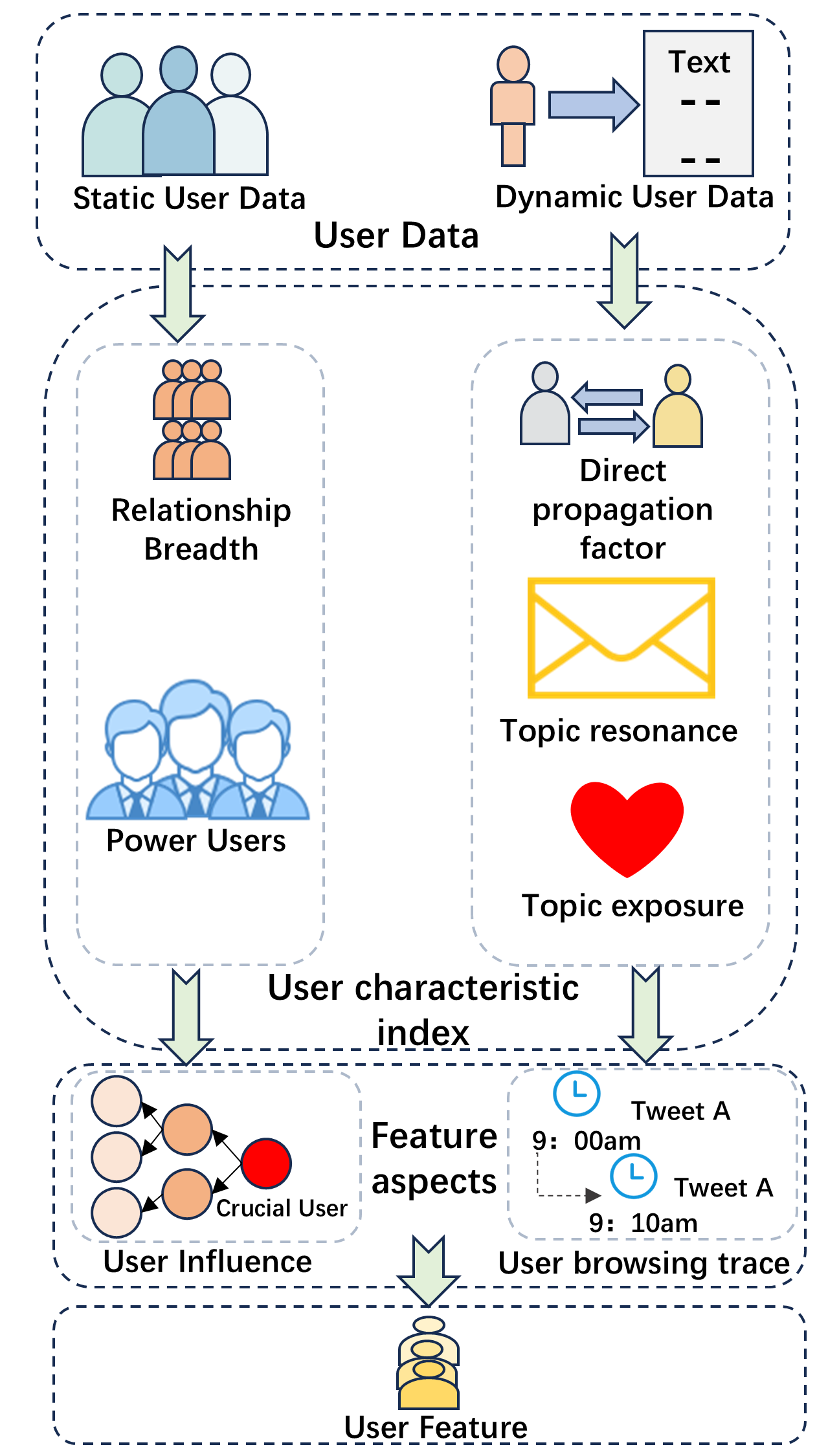}
   \caption*{Fig.2.Processing of User Feature}
\end{figure}

\subsection{Emotional Disposition}
Human behavior is influenced not only by rational considerations but also by emotional factors, particularly in the context of online information. Different demographic or social groups may exhibit significantly varied emotional responses to the same content, which in turn shapes their reactions and behaviors. For instance, the likelihood of retweeting increases when topic content evokes emotional resonance. Understanding the emotional tendencies embedded in topic content can thus provide critical insights into the direction and dynamics of topic dissemination, enabling more accurate predictions of its propagation trends.

To this end, this study examines the emotional factors inherent in the topic communication process and proposes an emotional tendency model to quantify their impact. The associated computational frameworks are defined in Eq. (8), Eq. (9), and Eq. (10).
\begin{equation}\label{eq8}
   Score\left( T_1,T_2,T_3...T_n \right) =Soft\max \left( y \right) =|y_{pos},y_{neg}|
\end{equation}

\begin{equation}\label{eq9}
   \varDelta =y_{pos}-y_{neg}
\end{equation}

\begin{equation}\label{eq10}
   M\left( T_i \right) =\begin{cases}
      Positive&		if\ \varDelta >0\\
      Negative&		if\ \varDelta <0\\
   \end{cases}
\end{equation}

$M\left( T_i \right)$ represents the sentiment tendency of the topic text $T_i$.$Score\left( T_1,T_2,T_3.....T_i \right)$ denotes the probability that text $T_i$ has the sentiment direction of the text derived from the sentiment classification model.$y_{pos}$    ref-ers  to the probability that the text belongs to the positive sentiment direction and $y_{neg}$ denotes the probability that the text belongs to the negative sentiment direction.

\subsection{Timing Characteristics}
Building on the intrinsic nature of topic dissemination, this study models the propagation of topic-related text as a time-series problem and systematically extracts its temporal features. By leveraging time-series computations (as defined in Eq. (11)), this approach effectively addresses the challenge of long-term dependencies in topic propagation while integrating contextual semantic information.

The extraction of time-series features enables a more comprehensive representation of temporal trends in the data, thereby enhancing the predictive capability for information dissemination. This method overcomes the limitations of traditional topic propagation models, which primarily emphasize user attributes and quantitative metrics, by incorporating temporal dynamics and contextual semantics. As a result, it significantly improves the accuracy of topic dissemination predictions.
\begin{equation}\label{eq11}
   h_t=\left[ \begin{matrix}
      h^f_t&		h^b\\
   \end{matrix}_t \right] 
\end{equation}
where $h_t$ is a temporal feature, $h^f_t$ as well as $h^b_t$ denote the hidden state vectors of the text at moment t, which are obtained by time series prediction modeling.The final temporal feature vector is obtained by splicing the hidden state vectors.
\section{Model Methods}
\subsection{Topic spread classification model based on fusion features}
In this study, a novel topic propagation classification model, MPT-PropNet (Multi-modal Propagation Transformer Propagation Network), is proposed based on the convergence characteristics of the topic propagation process.The model integrates four key dimensions—user attributes, emotional dynamics, interaction frequency, and temporal information—to enhance the generalization and realism of topic propagation classifications.Further Secure Management of the Social Network Information Environment.

Compared with other topic classification models, this paper's model has the following advantages:

a) Multi-dimensional feature integration: By incorporating diverse features, the model captures topic dissemination trends from multiple perspectives, thereby improving the accuracy and generalizability of classifications.

b) New user-based metrics: Two novel indices, user relationship breadth, and user authority, are introduced to more precisely quantify user influence within the propagation process.

c) User browsing traces: The model replaces traditional topic dissemination traces with user browsing traces, enabling a more effective quantification of dissemination trends and enhancing prediction accuracy regarding dissemination volume.

d) Time-series analysis: Unlike many existing studies, the model considers topic propagation as a time-series problem, leveraging temporal features to better predict dissemination patterns.

e) Advanced feature fusion: Drawing on insights from the Transformer architecture, a novel feature fusion method is introduced, significantly improving the effectiveness of feature integration and enhancing overall model performance.

The overall architecture of the model is as \textbf{Fig.3}:

\begin{figure*}[ht]
   \centering
   %\hspace{3500mm}
   \includegraphics[height=8.8cm,width=15cm,scale=1.5]{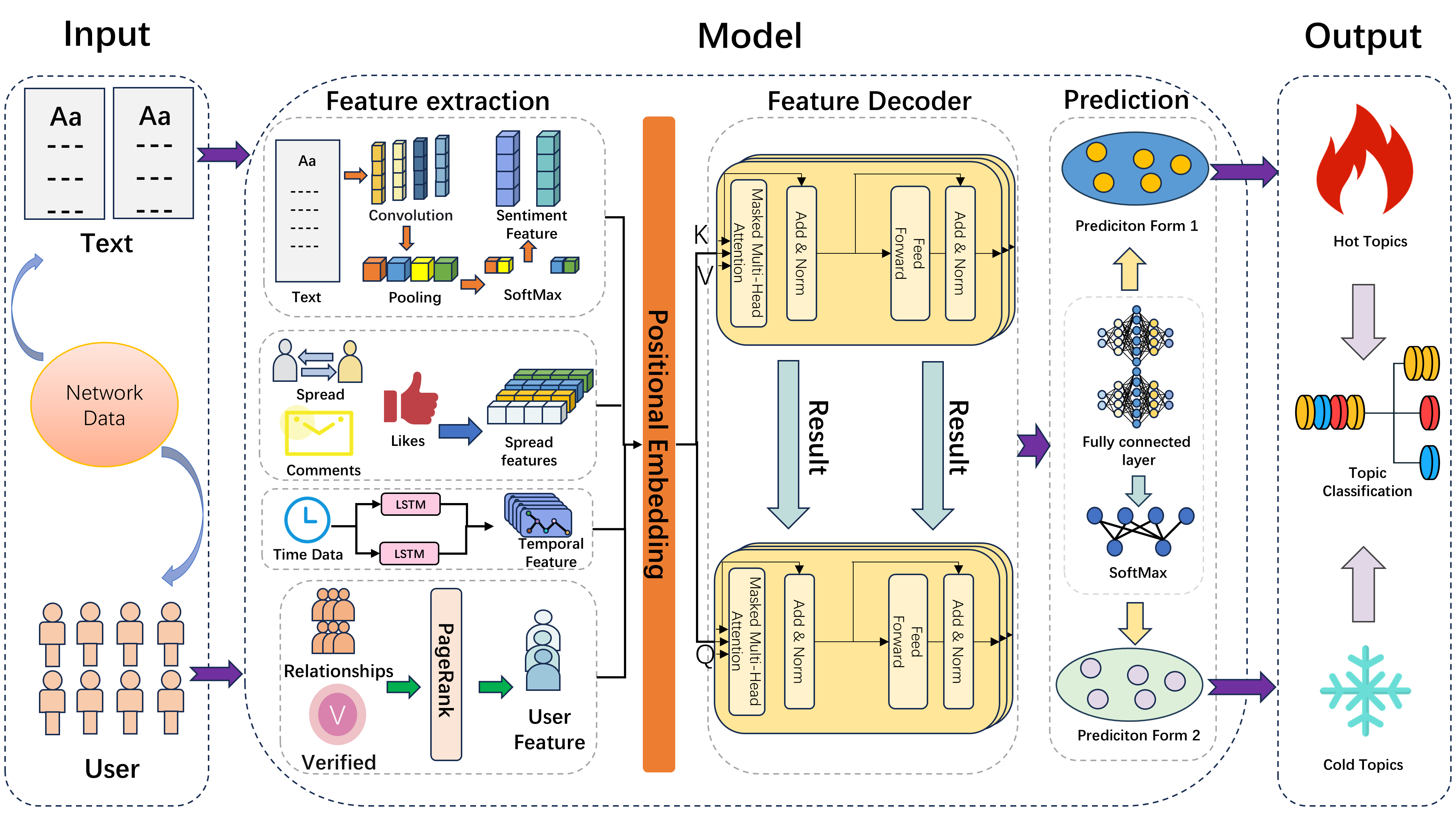}
   \caption*{Fig.3.Overview of MPT-PropNet for social emotion prediction to online trending topics.}
\end{figure*}
\subsection{Model Framework}
\subsubsection{UIR Model}
This study proposes a comprehensive evaluation model for user influence, termed the User-Influence-Ranking (UIR) model. This model is built upon the foundational principles of the PageRank algorithm and incorporates two novel metrics: User Relationship  Brea-dth, which quantifies the diversity and extent of a user’s connections within the network, and User Authority, which measures a user's credibility and impact. These metrics enable a more accurate assessment of user influence in social networks.
By analyzing the PageRank values, the model identifies key nodes within the network, whose characteristics are leveraged to capture and model the propagation dynamics of social networks. The mathematical formulation for the UIR model is provided in Eq. (12).
%\begin{equation}\label{eq12}
    %PR_{UIR}(U_i) = PR_{influence}(U_i) + d\sum_{U_j \in N(U_i)}\frac{PR_{UIR}(U_j)}{LR(U_j)}+1 - d
%\end{equation}

\begin{align}
   PR_{UIR}(U_i) &= PR_{influence}(U_i) + 
   d\sum_{U_j \in N(U_i)}\frac{PR_{UIR}(U_j)}{LR(U_j)} \notag \\
   &\quad + 1 - d
\end{align}

The $PR_{UIR}\text{(}U_i\text{)}$ represents the final PageRank valu  e of user $U_i$.$PR_{influence}\left( U_i \right) $ denotes the user's self-influence,which refers to Section 3.1 for details on the specific calculation methodologies used by various companies. $N\left( U_i \right) $ denotes the set of all users linking to the user, and $LR\left( U_j \right) $ denotes the number of external relationships of $U_j$, which can be expressed as the number of user follows.

Users' influence ranking is determined by calculating their PageRank values, where d represents the damping coefficient, typically set to d=0.85.

Using the influence rankings derived from the UIR model, user browsing traces within the topic information are assigned and quantified. This process enables forming the final user characteristic profiles associated with the topic information, providing a more precise representation of user influence within the dissemination dynamics.

\subsubsection{Sentiment Feature Extraction Model}
To extract sentiment features, this study employs the Text-CNN model for sentiment classification. Text-CNN is a convolutional neural network-based model specifically designed for text classification tasks. Its architecture consists of an embedding layer, convolutional layer, pooling layer, fully connected layer, and a SoftMax layer for classification.

Given an input text $x_i$,the embedding layer transforms it into a sequence of vectors, producing the text sequence vector $\boldsymbol{X}$.This vector is then passed through the convolutional layer,where the feature map $h_i$ is generated.Subsequently,the pooling layer extracts the most significant feature,denoted as $\boldsymbol{\hat{h}_i}$,which is used for downstream processing.

\begin{equation}\label{eq13}
   h_i = f\left( \sum_{x=1}^3 \sum_{y=1}^3 w_{(x,y)} c_{(x,y)} + b_i \right)
\end{equation}

\begin{equation}\label{eq14}
   \boldsymbol{\hat{h}_i} = \max \left\{ h_1, h_2, \ldots, h_n \right\}
\end{equation}

where $w_{(x,y)}$ denotes the corresponding weight matrix, $c_{(x,y)}$ is the input matrix, and $b_i$ denotes the offset matrix.

Finally, the feature vector $\boldsymbol{h}$ is obtained from the fully-connected layer and the final probability distribution of the sentiment category $y$ is obtained from the output layer, $W_h$ is the weight matrix of the fully-connected layer and $b_h$ is the bias vector:

\begin{equation}\label{eq15}
   \boldsymbol{h} = \left[ \boldsymbol{\hat{h}_1}, \boldsymbol{\hat{h}_2}, \ldots, \boldsymbol{\hat{h}_n} \right]
\end{equation}

\begin{equation}\label{eq16}
   y=soft\max \left( W_h\cdot \boldsymbol{h}+b_h \right) 
\end{equation}
\subsubsection{Temporal Feature Extraction Model}
To address the challenge of temporal feature extraction, this study employs the Bi-LSTM model (Bidirectional Long Short-Term Memory). The Bi-LSTM model enhances contextual understanding by extracting temporal features in both forward and backward directions, thereby capturing bidirectional dependencies. Furthermore, it effectively resolves the issue of long-term dependencies often encountered in sequential data processing.

The model calculates the hidden states $h^f_t$ (forward direction) and $h^b_t$ (backward direction) using Eq. (17) and Eq. (18), respectively. These hidden states are subsequently used as inputs for the next time step, ensuring the propagation of context-rich information through the sequence.
\begin{equation}\label{eq17}
   h^ft=LSTM\left( h^f_{t-1},x_t \right) 
\end{equation}

\begin{equation}\label{eq18}
   h^b_t=LSTM\left( h^b_{t-1},x_t \right)
\end{equation}

Finally, the hidden states are spliced to obtain the final timing feature $h_t$
\begin{equation}\label{eq19}
   h_t=\left[ 
      \begin{matrix}
       h^f_t&		 h^b\\
      \end{matrix}_t 
      \right] 
\end{equation}
\subsubsection{Fusion of multidimensional features}

Building upon the structure and principles of the Transformer model, this study introduces a novel feature fusion method to improve the integration of multi-dimensional features in the context of information dissemination. Unlike the traditional approach of simple linear summation fusion, the proposed method utilizes the Transformer’s architecture to achieve a more effective fusion of features.

In this study, the fused features are processed throug  h the fully connected layer and the SoftMax layer within the Transformer model to compute the probability of information dissemination,The formula is shown in Equation 20. This probability serves as the basis for determining whether a given topic is likely to achieve widespread dissemination. By leveraging the advanced architecture of the Transformer model, this approach ensures a more accurate and reliable prediction of dissemination trends.

\begin{equation}\label{eq20}
   Y_i=SoftMax\left( \text{Re}lu\left( W_3X+b_3 \right) W_4+b_4 \right) 
\end{equation}

where $Y_i$ denotes the probability of the message being propagated and $W_3,W_4,b_3,b_4$ are all learnable parameters

The Transformer model consists of three main components: the encoder, the decoder, and the embedding/output layers. Its overall architecture follows a sequence-to-sequence (seq2seq) framework, with the core mechanism being the attention mechanism. This mechanism includes multi-head self-attention, multi-head mas  ked self-attention, and multi-head cross-attention, which collectively enable the model to capture intricate dependencies within and across sequences.
Embedded input data is processed through the attention mechanism to extract contextual relationships. The resulting representations are passed through the output layer, comprising a fully connected layer and a SoftMax layer, to calculate the probability distribution over the output vocabulary. The word with the highest probability is selected as the final prediction.

%\begin{figure*}[h]
   %\centering
   %\hspace{3500mm}
   %\includegraphics[height=8cm,width=8cm,scale=1.5]{transformer.png}
   %\caption*{Fig.3.Structure of Transformer}
%\end{figure*}

To represent the sequence of topics being forwarded, positional embedding is applied to the features of topics, resulting in enhanced topic representations. These enriched features are then fed into the Decoder layer of the Transformer model, where they are assigned as the Key, Query, and Value sets for the attention mechanism. This approach ensures a comprehensive and efficient fusion of features.
The specific formulation is provided as follows:

\begin{equation}\label{eq21}
   Attention\left( K,Q,V \right) =SoftMax\left( \frac{QK^T}{\sqrt{d_k}}+C \right) V
\end{equation}

\begin{equation}\label{eq22}
   a_{i\text{,}k}=Attention\left( T_{i}^{'}W_{k}^{Q},H_{i}^{'}W_{k}^{K},H_{i}^{'}W_{k}^{V} \right) 
\end{equation}

\begin{equation}\label{eq23}
   A_i=\left[ \begin{matrix}{l}
      a_{i\text{,}1}&		a_{i,2}&		...&		a_{i,head-1}&		a_{i,head}\\
   \end{matrix} \right] W^Q
\end{equation}

In Eq. (22),$W_{k}^{Q},W_{k}^{V},W_{k}^{K}$ as well as $W^Q$ are learnable weight parameters, and specific values can be obtained through model training.$d_k=d/head$, $head$ denotes the number of heads of multi-head attention, $C$ is used to mask the weights of future timestamps, ensuring that the prediction of position $j$ can only rely on sequences smaller than $j$, and $A_i$ denotes the fusion feature of message $i$.

\section{Experiments}
\subsection{Experiment Setup}
\subsubsection{Datasets}
To comprehensively analyze the characteristics and dynamics of topic dissemination in social networks, this study utilizes Tweet datas from Twitter, a globally oriented social platform, as experimental data. Specifically, data were collected from three trending topics on Twitter between July and September 2022: \#iphone14\# \footnote{\url{https://www.kaggle.com/datasets/tleonel/iphone14-tweets}}, \#Cost of Living\# \footnote[2]{\url{https://www.kaggle.com/datasets/tleonel/cost-of-living}} and \#crypto\#\footnote[3]{\url{https://www.kaggle.com/datasets/tleonel/crypto-tweets-80k-in-eng-aug-2022}}.

Due to the absence of sentiment annotations and ground truth predictive values in the original dataset, sentiment labels were assigned to each Tweet based on its textual content. Each Tweet was categorized into one of three sentiment classes: 'positive', 'neutral', or 'negative'. Additionally, a prediction label for each Tweet was established by incorporating various features: the number of likes, retweets, and comments; the number of user followers; user verification status; and time-related factors (e.g., Tweets posted during the daytime were given additional weight). Using a weight assignment method and a predetermined prediction threshold, Tweets were scored. If a Tweet’s score exceeded the threshold, it was labeled as propagated; otherwise, it was labeled as non-propagated. This process facilitated the binary classification required for prediction labeling.

Given the complexity of the Tweet dataset, preprocessing was performed to clean and standardize the textual content. Irrelevant features such as user location and profile descriptions were removed. Noise from emoticons, punctuation, hyperlinks, and user references was eliminated to ensure greater focus on the textual content itself. These steps were undertaken to improve the accuracy and efficiency of the experimental analysis.

The processed dataset’s key characteristics are summarized in Table 1, where Users denotes the number of users in the dataset, Relations represents the average correlation of user node social relationships, and Volume indicates the number of propagation sequences for a single topic sorted by time.

Due to the influence of environmental and other factors, this paper intercepts a portion of the data for experiments.Detailed information on the data is provided in Table 1.

\begin{table*}[ht]
   \centering
   \caption*{Table.1.Partial Data Containing Elements}
   \setlength{\tabcolsep}{1.3cm}{
      \begin{tabular}[c]{cccc}
         \toprule
         Topics & Users & Relations & Volume \\
         \midrule[2pt]
         \#Iphone14\#  &  26,638 &  1,611 & 3,543 \\
         \#Cost of Living\#   &  29,990  & 1,966 & 4,168 \\
        \#Crypto\#  &  17,832     &  1,849 & 3,762 \\
         \bottomrule
      \end{tabular}

   }
\end{table*}

Following the methodology of most previous studies, this paper splits the data for each topic into three subsets: 80\% for training, 10\% for validation, and 10\% for testing.

\subsubsection{Basline method}
In the context of topic propagation prediction, existing research generally approaches the problem through two types of task frameworks: classification tasks and time-series prediction tasks. This study specifically focuses on the binary classification task of determining whether a topic will propagate. Building on insights from prior research and empirical experimentation, we systematically evaluate the performance of different models across three paradigms: traditional machine learning, unimodal deep learning, and multimodal approaches, including graph-based techniques. Specifically, we employ Logistic Regression to represent machine learning methods, Text-CNN to capture the performance of unimodal deep learning models, and GNN (Graph Neural Network) to assess graph-based neural networks. The distinction between machine learning and deep learning is illustrated through a comparative analysis of Logistic Regression and Text-CNN. To investigate the impact of modality, we compare the performance of the unimodal Text-CNN with the multimodal MPT-PropNet model. Finally, we analyze the strengths and limitations of graph-based models and multimodal approaches by evaluating GNN and MPT-PropNet. This comprehensive comparison provides insights into the effectiveness and trade-offs of these meth  odologies in addressing the topic propagation prediction problem.

1) Logistic Regression (LR)\Citep{BokoloandLiu2024}.   Logistic Regression is a fundamental linear classification model in machine learning that predicts the probability of topic propagation within a social network. It operates by learning a linear relationship between input features and target categories. The core principle of the model is to transform the linear regression output into probability values using the logistic function, ensuring the predictions fall within the range [0, 1].

2) GNN\citep{Shan2024}. Graph Neural Networks (GNNs) have emerged as powerful tools for learning from graph-s   tructured data, making them particularly suitable for tasks like topic propagation prediction in social networks. Unlike traditional linear models such as Logistic Regression, which maps input features to probabilities using a logistic function, GNNs extend this concept by incorporating relational information from graph structures.GNNs aggregate information from neighboring nodes and edges within a graph, enabling the model to capture complex, non-linear relationships and dependencies among interconnected entities. By iteratively propagating and updating node representations, GNNs generalize the concept of feature transformation and probability prediction to graph-based data, allowing for more accurate modeling of propagation phenomena in social networks.

3) Text-CNN\citep{Liu2024}. Text-CNN is a classical deep-learning model specifically designed for processing text data. It excels at extracting local contextual features by applying convolutional operations over word embeddings, making it highly effective for tasks involving topic-related text content. Due to its ability to capture n-gram level features and maintain computational efficiency, Text-CNN has been widely adopted for various NLP applications, including text classification, sentiment analysis, and topic modeling. Furthermore, its architecture is well-suited for multi-task learning, enabling simultaneous optimization across related tasks by leveraging shared feature representations.

\subsection{Performance Evaluation}
\subsubsection{Experimental Analysis}
Experiments were conducted on three distinct topic datasets to evaluate the performance of the model on the propagation binary classification task. The performance was assessed by examining the model's ability to predict both positive and negative samples accurately across different datasets. A confusion matrix heatmap was utilized to provide a clearer visualization of the model's prediction capabilities across various categories. This heatmap highlights the classification accuracy for positive and negative samples, offering detailed insights into the model's performance. The results are illustrated in \textbf{Fig.4},  \textbf{Fig.5} and \textbf{Fig.6}.
%\begin{figure*}[ht]
   %\centering
   %\begin{subfigure}{0.45\textwidth}
       %\centering
       %\includegraphics[width=\textwidth]{martix_costofliving.png}
       %\label{fig:model1}
   %\end{subfigure}
   %\hfill
   %\begin{subfigure}{0.45\textwidth}
       %\centering
       %\includegraphics[width=\textwidth]{martix_crypto.png}
       %\label{fig:model2}
   %\end{subfigure}
   
   %\vspace{1em} % 调整上下子图之间的间距
   
   %\begin{subfigure}{0.5\textwidth}
       %\centering
       %\includegraphics[width=\textwidth]{martix_iphone.png}
       %\label{fig:model3}
   %\end{subfigure}
   
   %\caption*{Fig.4 Confusion matrices of MPT-PropNet on different topics.}
   %\label{fig:confusion_matrices}
%\end{figure*}

\begin{figure}[ht]
   \centering
   %\hspace{30mm}
   \includegraphics[height=7cm,width=8cm,scale=1.5]{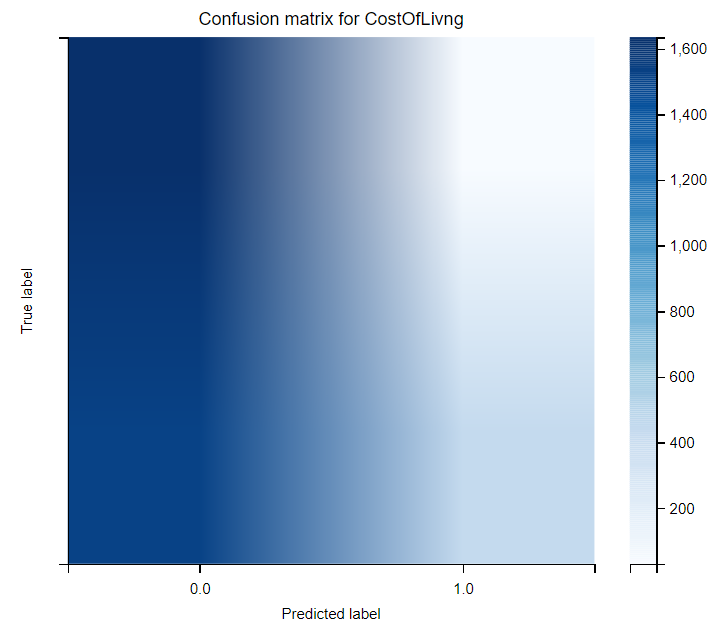}
   \caption*{Fig.4.Confusion matrices of MPT-PropNet on CostOfLiving.}
   
\end{figure}

\begin{figure}[ht]
   \centering
   %\hspace{30mm}
   \includegraphics[height=7cm,width=8cm,scale=1.5]{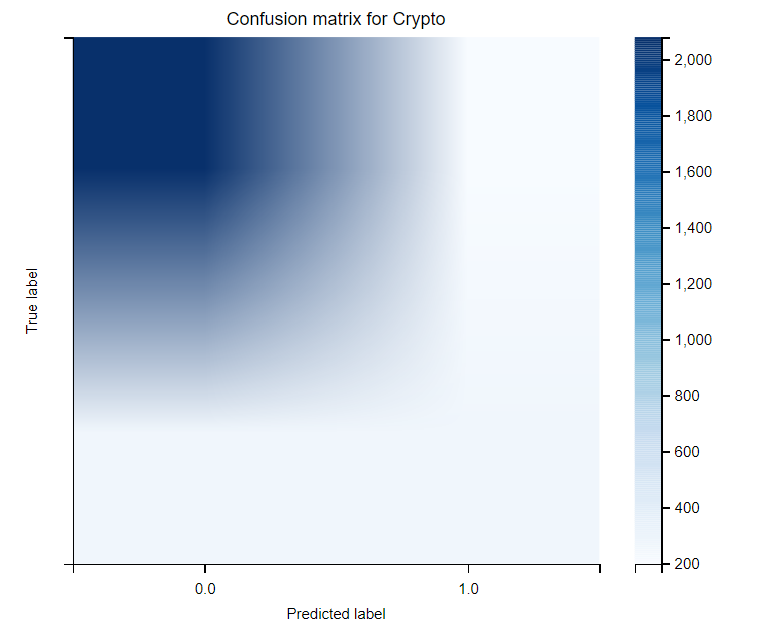}
   \caption*{Fig.5.Confusion matrices of MPT-PropNet on Crypto.}
   
\end{figure}

\begin{figure}[ht]
   \centering
   %\hspace{30mm}
   \includegraphics[height=7cm,width=8cm,scale=1.5]{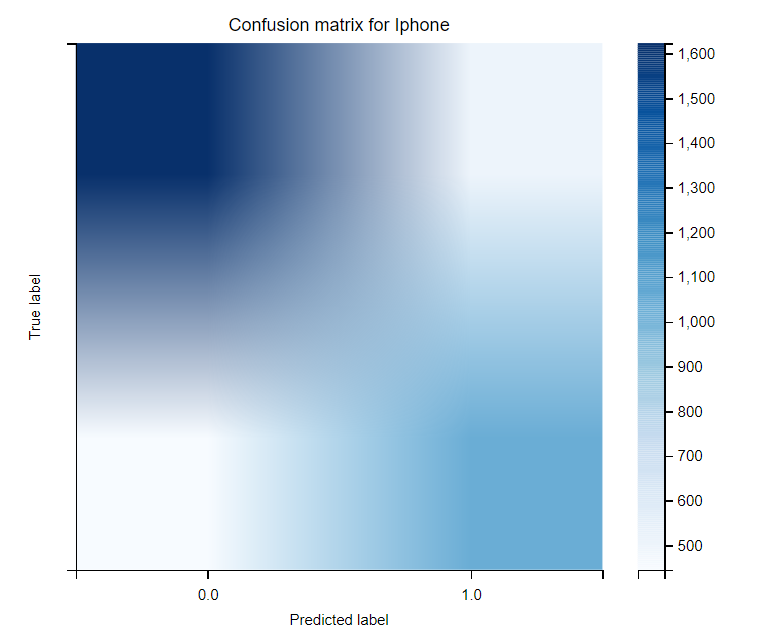}
   \caption*{Fig.6.Confusion matrices of MPT-PropNet on Iphone.}
   
\end{figure}

As illustrated in \textbf{Fig.4},   \textbf{Fig.5} and \textbf{Fig.6}, MPT-PropNet demonstrates strong performance across all three datasets. The confusion matrix heatmap clearly shows that the intensity of the color in the upper-left corner of the map is significantly darker than the other regions. In contrast, the lower-right corner is notably lighter, particularly in the \#Crypto\# dataset, where the color approaches near transparency. This pattern indicates that the model achieves many True Positives (TP) and True Negatives (TN) across different datasets, successfully classifying samples in most cases. The lighter color in the lower-right corner suggests a relatively low number of False Positives (FP) and False Negatives (FN), highlighting the model's strong ability to distinguish between positive and negative samples. Consequently, MPT-PropNet exhibits high accuracy with a low misclassification rate in classifying both positive and negative samples. Overall, the model performs excellently across all datasets, underscoring its effectiveness and superiority in predicting topic propagation.

\subsubsection{Model Comparison Effectiveness}

To validate the accuracy and generalizability of the proposed MPT-PropNet in topic prediction, three evaluation metrics are employed: AUC, F1 Score, and Recall. The AUC  quantifies the performance of the classification model by measuring the trade-off between the true positive rate and the false positive rate. The F1 Score assesses the balance between precision and recall, reflecting the model's ability to interpret the influence of features on its output effectively. Finally, Recall represents the proportion of correctly identified positive samples relative to the total actual positive samples, providing insight into the model's ability to predict topic propagation accurately.
Higher values for these metrics, especially those approaching 1, indicate superior performance of the topic propagation prediction model. The corresponding formulae are provided in Eq. (24), Eq. (25), and Eq. (26).

\begin{equation}\label{eq24}
   Recall=\frac{TP}{TP+FN}
\end{equation}

where TP denotes the true example and FN denotes the false negative example.
\begin{equation}\label{eq25}
   AUC=\frac{1}{2}\sum_{i=1}^{n-1}\left( x_{i+1}-x_i \right)\left( y_i+y_{i+1} \right)
\end{equation}

where $x_i$ denotes the false positive rate and $y_i$ denotes the true rate.
\begin{equation}\label{eq26}
   FI=\frac{2\times Precision\times Recall}{Precision+Recall}
\end{equation}

\begin{equation}\label{eq27}
   Precision=\frac{TP}{TP+FP}
\end{equation}

where Precision denotes the rate of precision,  and FP denotes false positive examples.

We compare the effects on the evaluation metrics of the traditional machine learning Logistic Regression model, the unimodal deep learning model Text-CNN, and the graphical convolutional neural network model GNN, as shown in Table.2

\renewcommand{\arraystretch}{2.5}
\begin{table*}[ht] 
    \centering
    \caption*{Table.2. Prediction of public opinion features under different combinations of attributes.}
    \resizebox{0.8\textwidth}{!}{
    \begin{tabular}{l|ccc|ccc|ccc}
       \toprule
       \multirow{2}{*}{\textbf{Modules}} & \multicolumn{3}{c|}{\textbf{\#iphone14\#}} & \multicolumn{3}{c|}{\textbf{\#Cost of Living\#}} & \multicolumn{3}{c}{\textbf{\#crypto\#}} \\
       \cmidrule(lr){2-4} \cmidrule(lr){5-7} \cmidrule(lr){8-10}
       & Recall & AUC & FI & Recall & AUC & FI & Recall & AUC & FI \\
       \midrule
       LR                  & 0.570 & 0.602 & 0.574 & 0.459 & 0.608 & 0.289 & 0.812 & 0.736 & 0.760 \\
       Text-CNN                  & 0.589 & 0.534 & 0.437 & 0.459 & 0.508 & 0.290 & 0.813 & 0.565 & 0.729 \\
       GNN                  & 0.719 & 0.719 & 0.702 & 0.601 & 0.601 & 0.506 & 0.686 & 0.686 & 0.824 \\
       MPT-PropNet                  & \textbf{0.736} & \textbf{0.730} & \textbf{0.737} & \textbf{0.632}& \textbf{0.629} & \textbf{0.546} & \textbf{0.835} & \textbf{0.812} & \textbf{0.825} \\
       \bottomrule
    \end{tabular}}
\end{table*}

The results indicate that MPT-PropNet consistently outperforms all other models across all datasets, achieving the highest Recall, AUC, and F1 scores, thereby demonstrating its robustness and superior predictive capability. Although GNN performs better than both Text-CNN and LR, MPT-PropNet’s ability to adapt to a diverse range of datasets, including the challenging \#Cost of Living\# dataset, underscores its effectiveness in capturing the complexities inherent in public opinion dynamics. In contrast, LR and Text-CNN exhibit relatively lower performance, particularly on datasets that require more nuanced feature extraction. In summary, MPT-PropNet consistently delivers superior results, emphasizing its potential for broader application in opinion prediction tasks and social network analysis.

To further elucidate the relative strengths and weaknesses of the various models, we plotted the ROC curves for each model across the three topics. These curves were then optimized using cubic spline interpolation, resulting in smoother final ROC curves for each model on the corresponding datasets, as depicted in \textbf{Fig.7},  \textbf{Fig.8} and \textbf{Fig.9}. From the figures, it is evident that the ROC curves of the MPT-PropNet model consistently lie above those of the other models across all three topics, highlighting its superior predictive performance across diverse topic areas.

%\begin{figure*}[ht]
   %\centering
   %\begin{subfigure}{0.45\textwidth}
       %\centering
       %\includegraphics[width=\textwidth]{ROC_iphone1.png}
       %\caption{\#iphone14\#}
       %\label{fig:model1}
   %\end{subfigure}
   %\hfill
   %\begin{subfigure}{0.45\textwidth}
       %\centering
       %\includegraphics[width=\textwidth]{ROC_costoflivng1.png}
       %\caption{\#Cost of Living\#}
       %\label{fig:model2}
   %\end{subfigure}
   
   %\vspace{1em} % 调整上下子图之间的间距
   
   %\begin{subfigure}{0.5\textwidth}
       %\centering
       %\includegraphics[width=\textwidth]{ROC_alcrypto1.png}
       %\caption{\#crypto\#}
       %\label{fig:model3}
   %\end{subfigure}
   
   %\caption*{Fig.6 ROC curves of different models on corresponding datasets}
   %\label{fig:confusion_matrices}
%\end{figure*}

\begin{figure}[ht]
   \centering
   %\hspace{30mm}
   \includegraphics[height=7cm,width=8cm,scale=1.5]{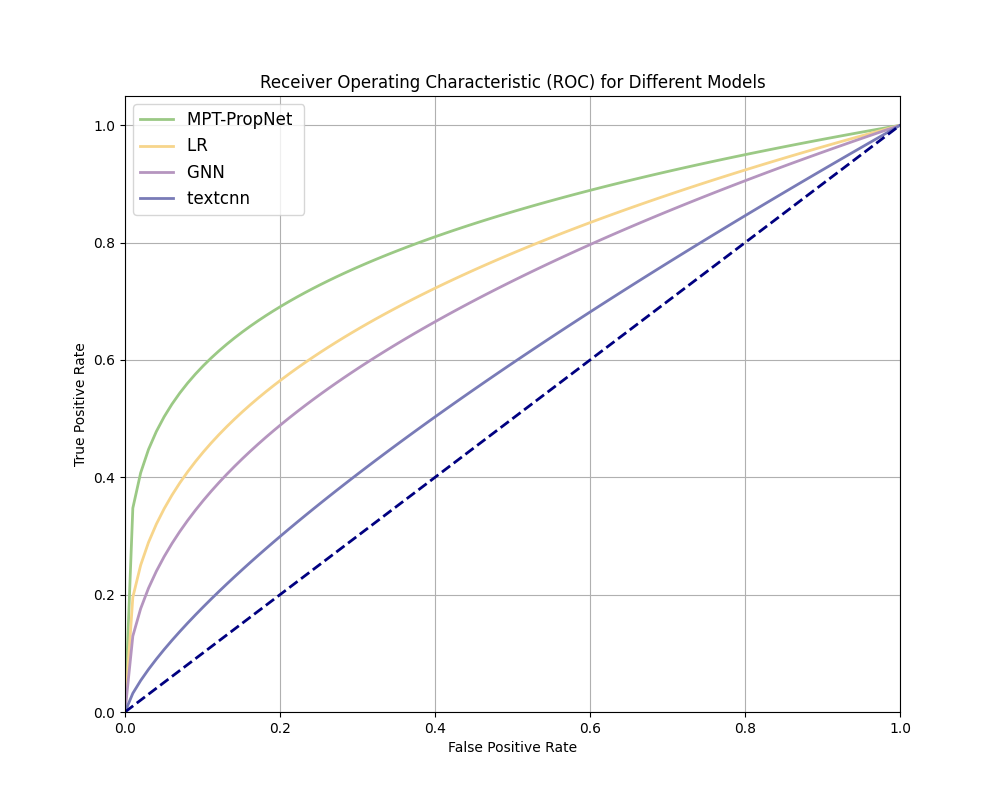}
   \caption*{Fig.7.ROC curves of different models on \#crypto\#}
   
\end{figure}

\begin{figure}[ht]
   \centering
   %\hspace{30mm}
   \includegraphics[height=7cm,width=8cm,scale=1.5]{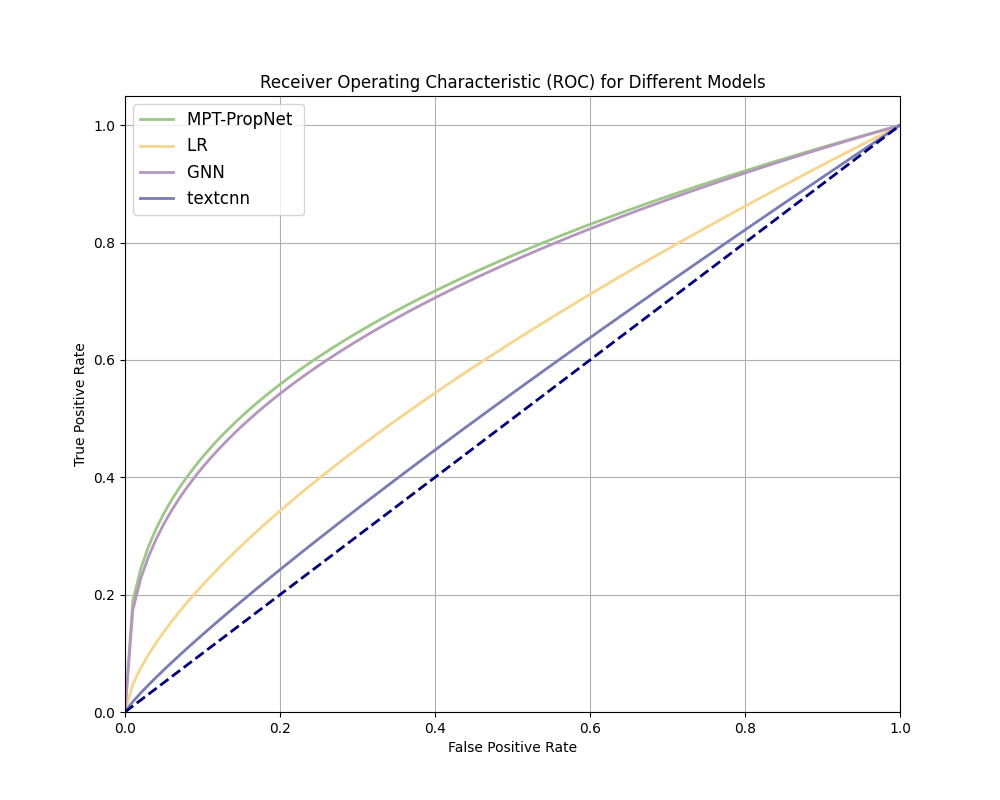}
   \caption*{Fig.8.ROC curves of different models on \#iphone14\#.}
   
\end{figure}

\begin{figure}[ht]
   \centering
   %\hspace{30mm}
   \includegraphics[height=7cm,width=8cm,scale=1.5]{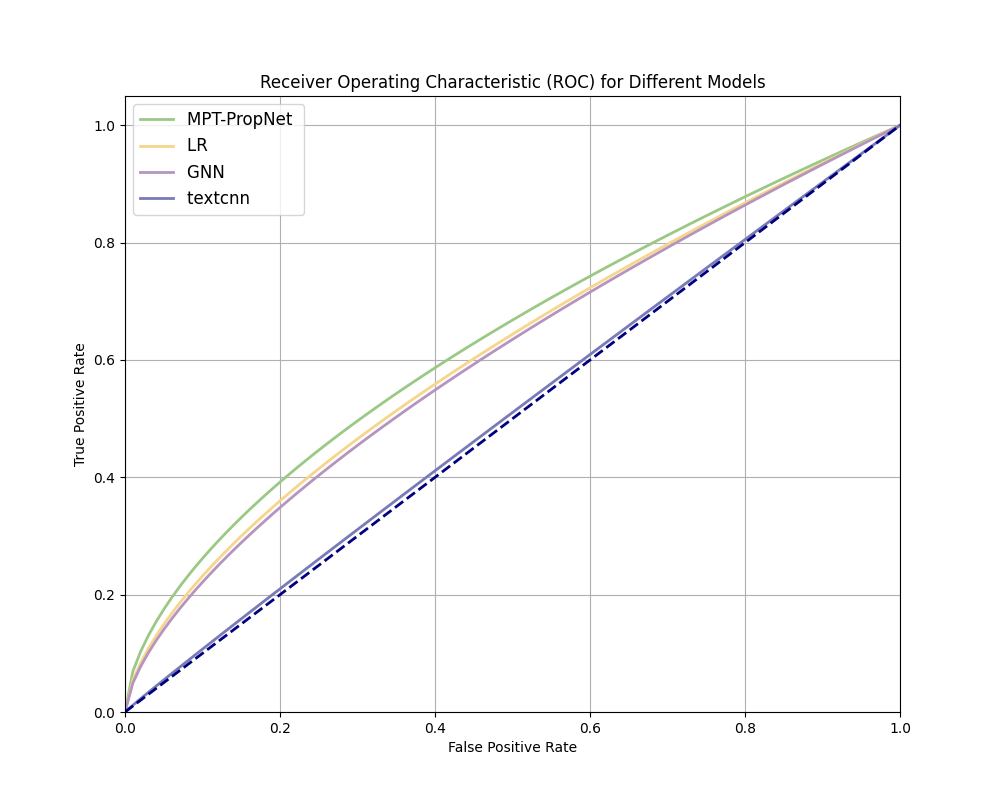}
   \caption*{Fig.9.ROC curves of different models on \#Cost Of Living\#.}
   
\end{figure}

\begin{figure*}[ht]
   \centering
   %\hspace{3500mm}
   \includegraphics[height=7cm,width=10cm,scale=1.5]{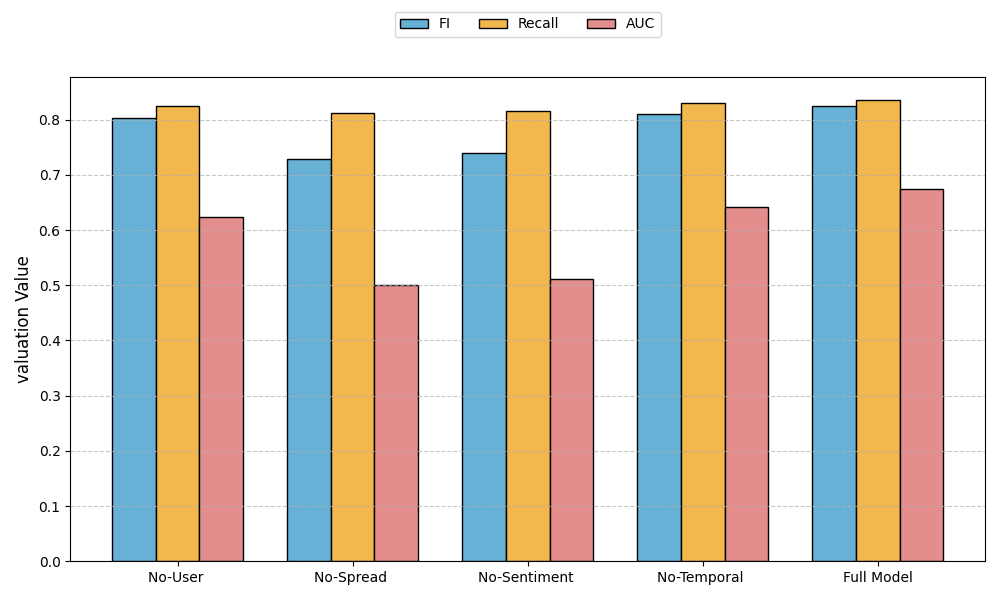}
   \caption*{Fig.10. MPT-PropNet remove modules}
\end{figure*}

As illustrated in \textbf{Fig.7},  \textbf{Fig.8},  \textbf{Fig.9} and Table 2, the MPT-PropNet model demonstrates performance co-mparable to the GNN model in terms of the AUC value for the \#iphone14\# and \#CostOfLiving\# topics. This is primarily attributed to the fact that the GNN model constructs a social network topology, which aligns more closely with the structure of user relationships, making it more suitable for capturing user interaction dynamics than the UIR model proposed in this study. However, despite this advantage, the GNN model's performance is hindered by its overemphasis on user features and its relatively reduced attention to other critical modules, leading to suboptimal overall performance.

In contrast, the Logistic Regression (LR) model, which is a machine learning model focused on identifying linear relationships, performs poorly in this context. This is because the relationships in social networks are often non-linear and highly complex, a limitation that LR struggles to address, making it the least effective model among those tested.

On the other hand, the MPT-PropNet model, due to its sophisticated network structure and enhanced feature extraction capabilities, can better capture the intricate patterns and interdependencies within the data. This enables the model to more effectively predict the propagation characteristics and public opinion dynamics across various topics. The ROC curves for all three topics consistently outperform those of the other models, confirming the superior predictive performance of MPT-PropNet. This robust performance across different topics indicates that the MPT-PropNet model is better suited for handling the complex and diverse nature of topic propagation in social networks.

By combining these advanced components, the MPT-PropNet model demonstrates a significant performance improvement in topic propagation prediction, effectively addressing the limitations of existing approaches.

\subsection{Ablation Experiment}
Ablation experiments were conducted to comprehensively evaluate the contribution of each module in the proposed MPT-PropNet model to the overall prediction performance. These experiments systematically removed individual key modules to observe their impact on model behavior and performance metrics. Specifically, the following modules were analyzed: the user feature module, which captures user influence and relationship breadth; the emotion feature module, responsible for quantifying sentiment trends in topic dissemination; the timing feature module, designed to incorporate temporal dependencies; and the propagation feature module, which models the characteristics of information spread.
Each module was excluded from the model architecture independently, and the corresponding performance metrics, including AUC, F1-score, and Recall, were recorded to assess the contribution of the omitted module. By comparing the evaluation indexes before and after module removal, the importance of each component was quantitatively judged.

Since the primary goal of this experiment was to analyze the role of individual modules without introducing dataset-specific biases, the \#crypto\# dataset was selected for testing due to its balanced and representative nature. The performance degradation caused by the exclusion of each module highlights its significance in improving the model's predictive accuracy. The detailed results, illustrating the impact of module removal, are visualized in \textbf{Fig.10}.

The experimental results demonstrate that the removal of any one of the four modules—user features, communication features, sentiment features, or temporal features—leads to a significant decline in the model’s Recall, AUC, and FI scores. This indicates that each module makes a unique and indispensable contribution to the performance of the proposed model. By effectively integrating multi-dimensional features, the complete MPT-PropNet model achieves optimal performance through the efficient fusion of user attributes, communication patterns, emotional dynamics, and temporal information. These findings underscore the robustness and importance of the proposed framework, highlighting its significance in advancing topic propagation prediction in complex social networks.
\section{Discussion}
\subsection{Research Significance}
In this paper, we propose a multimodal feature fusion model, MPT-PropNet, which integrates user attributes, communication modes, emotional dynamics, and temporal information to achieve precise classification of topic dissemination. This approach facilitates more efficient public opinion monitoring, thereby enhancing the security and reliability of social network information. The MPT-PropNet model effectively captures public opinion features across different topics and demonstrates high recall, AUC, and F1-score. These performance metrics confirm its ability to accurately capture and predict public opinion features under various topics. The precise categorization of social network topic dissemination in this study enables timely insights into public opinion trends, real-time monitoring of social network information flow, and serves as a valuable reference for maintaining social stability and network security.

Although this study addresses key issues related to topic propagation in social networks, several technical and theoretical challenges remain to be explored. For instance, this paper considers only a subset of metrics related to user attributes, sentiment, dissemination patterns, and temporal sequences. However, social networks, as complex information systems, encompass a far broader range of metrics. Future research can refine and extend this study through more in-depth investigations into additional factors influencing topic propagation.
\subsection{Future Research Directions}
Future work will extend this study to additional information dissemination platforms and explore cross-platform dissemination patterns to enhance the model’s generalizability. Additionally, future research will expand the feature set of topic dissemination networks, particularly by examining the dynamic effects of user mood fluctuations and modeling dissemination characteristics across different time windows. Furthermore, we will investigate early warning mechanisms within the topic trend prediction model to improve sensitivity to emerging hotspots and enhance classification accuracy.
\section{Conclusion}
Understanding the evolution patterns and development trends of topic propagation is of great significance for public opinion monitoring, crisis prevention and control, early warning systems, and precision marketing. In this paper, we conduct an in-depth analysis of the influencing factors and their evolutionary dynamics during the topic dissemination process, while predicting topic development trends based on dissemination characteristics. The main contributions of this study include: the extraction and analysis of multimodal topic propagation features in social networks; the investigation of the evolution process of microblog topic propagation networks; and the prediction of topic propagation trends using a spatiotemporal attention model with multimodal fusion features.

\bibliography{paper}

\end{document}